\documentclass[superscriptaddress,twocolumn,showpacs,showkeys,preprintnumbers,amsmath,amssymb]{revtex4}
 
\usepackage{graphicx}
\usepackage{dcolumn}
\usepackage{bm}
 
\begin{document}

\preprint{APS/123-QED}

\title{Optimization in complex networks}
 
\author{Ramon Ferrer i Cancho}
\affiliation{Complex Systems Research Group, Department of Physics-UPC, Campus Nord, B4-B5, Barcelona 08034, SPAIN}
\author{Ricard V. Sol\'{e}}
\affiliation{Complex Systems Research Group, Department of Physics-UPC, Campus Nord, B4-B5, Barcelona 08034, SPAIN}
\affiliation{ICREA, Lluis Companys 23, Barcelona 08010, SPAIN}
\affiliation{Santa Fe Institute, 1399 Hyde Park Road, New Mexico 87501, USA.}

\date{\today}

\begin{abstract}
Many complex systems can be described in terms of networks 
of interacting units. Recent studies have 
shown that a wide class of both natural and artificial nets 
display a surprisingly widespread feature: the presence of highly
heterogeneous distributions of links, providing an extraordinary source 
of robustness against perturbations. Although most theories 
concerning the origin of these topologies use growing graphs, here we show 
that a simple optimization process can also account for 
the observed regularities displayed by most complex nets. 
Using an evolutionary algorithm involving minimization of link density 
and average distance, four major types of networks are encountered: 
(a) sparse exponential-like networks, (b) sparse scale-free networks, 
(c) star networks and (d) highly dense networks, apparently defining three major phases. These constraints provide a new explanation for scaling of exponent about $-3$. The evolutionary consequences of these results are outlined.
\end{abstract}


\maketitle

Many essential features displayed by complex systems, such 
as memory, stability and homeostasis emerge from 
their underlying network structure \cite{Strogatz2001,Kauffman1993}. 
Different networks exhibit different features at different levels. 
Although some nets, as those present in specific areas of the brain cortex, are 
densely wired \cite{Shepherd1990}
most complex networks are extremely sparse and exhibit the 
so-called small-world phenomenon \cite{Watts1998}. An inverse measure 
of sparseness, the so-called network density, is defined as 
\begin{equation}
\rho=\frac{\left< k \right>}{n-1}
\label{equation:density}
\end{equation} 
where $n$ is the number of vertices of the network and $\left<k \right>$ 
is its average degree. For real networks we have $\rho \in [ 10^{-5},10^{-1} ]$ 
\cite{ComplexNetworksDensityStatistics}.

It has been shown that a wide range of 
real networks can be described by a degree distribution 
$P(k) \sim  k^{-\gamma} \phi(k/\xi)$ where $\phi(k/\xi)$ introduces a cut-off 
at some characteristic scale $\xi$. 
Three main classes can be defined \cite{Amaral2000}. (a) When $\xi$ is very small, 
$P(k) \sim \phi(k/\xi)$ and thus the link distribution is single-scaled. 
Typically this would correspond to exponential or Gaussian distributions; 
(b) as $\xi$ grows, a power law with a sharp cut-off is obtained; (c) for 
large $\xi$, scale-free nets are observed. The last two cases have 
been shown to be widespread and their topological properties have immediate 
consequences for network robustness and fragility \cite{Barabasi2001a}. 
The three previous scenarios are observed in: (a) power grid systems 
and neural networks \cite{Amaral2000}, (b) protein interaction maps \cite{Jeong2001}, 
metabolic pathways \cite{Jeong2000} and electronic circuits \cite{Ferrer2001e} 
and (c) Internet topology \cite{Jeong2000}, 
scientific collaborations \cite{Newman2001a} and \cite{Ferrer2001a} lexical networks. 

\begin{figure}
\includegraphics[scale=0.3]{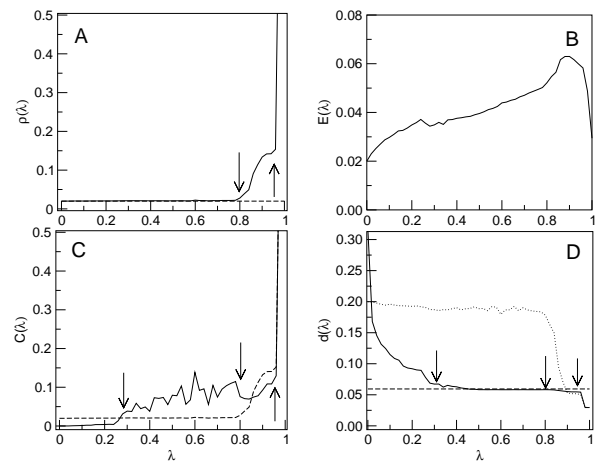}
\caption{\label{fig:properties_versus_lambda} 
Density (A), energy (B), clustering coefficient (C) and distance (D)
 as a function of $\lambda$.
Averages over $50$ optimized networks with $n=100$, 
$M={n \choose 2}$, $\mu=2/{n \choose 2}$ and $\rho(0)=0.2$ are shown.
 A: the optimal network becomes a clique for $\lambda$ close to $1$. 
The density of a star network, $\rho_{star}=2/n=0.02$ is shown as reference
 (dashed line). The clustering coefficient of a Poissonian network
 $C_{random}={\left< k \right>}/n$ is shown as reference in C. Notice
 that $C_{random} \approx \rho$. The normalized distance of a star
 network is, $d_{star}=6/(n+1)=0.059$ (dashed line) and that of a
 Poissonian network, $d_{random}={\log n}/{\log \left< k \right>}$ 
(dotted line) are shown as reference in D.}
\end{figure}

Scale-free nets are particularly relevant due to their extremely high homeostasis 
against random perturbations and fragility against removal of highly 
connected nodes\cite{Albert2000}. 
These observations have important consequences, from evolution to therapy \cite{Jeong2001}. 
One possible explanation for the origin of the observed distributions 
would be the presence of some (decentralized) optimization process. 

Network optimization is actually known to play a leading role in explaining allometric scaling in 
biology \cite{West1997,Brown2000,Banavar1999}: under the assumption of minimal cost in 
transportation, the scaling properties of branching structures are 
quite well understood as the outcome of an optimization process. 
In a related context, local and/or global optimization has been also
shown to provide remarkable result within the context of channel
networks \cite{Iturbe1997}. By using optimality criteria linking energy
dissipation and runoff production, the fractal properties in the
model channel nets were essentially indistinguishable from those observed in nature. 

Within the context of complex networks in which the only relevant elements are 
vertices and connections, it has been shown that several mechanisms of network 
evolution lead to scale-free structures \cite{Barabasi1999}. Optimization has not been 
found to be one of them. In a recent study, however, it was shown that
(Metropolis-based) minimization of both vertex-vertex distance and
link length (i.e. Euclidean distance between 
vertices)\cite{Mathias2001} can lead to the small-world
 phenomenon and hub formation. Nonetheless, this view takes into account 
Euclidean distance between vertices, which is generally not relevant in real 
complex networks. Here we show how minimizing both vertex-vertex distance and 
the number of links leads (under certain conditions) to the different
types of network topologies depending on the weight given to each
constraint. These two constraints include two relevant 
aspects of network performance: the cost of physical links between
units and communication speed among them. 

\begin{figure*}
\includegraphics{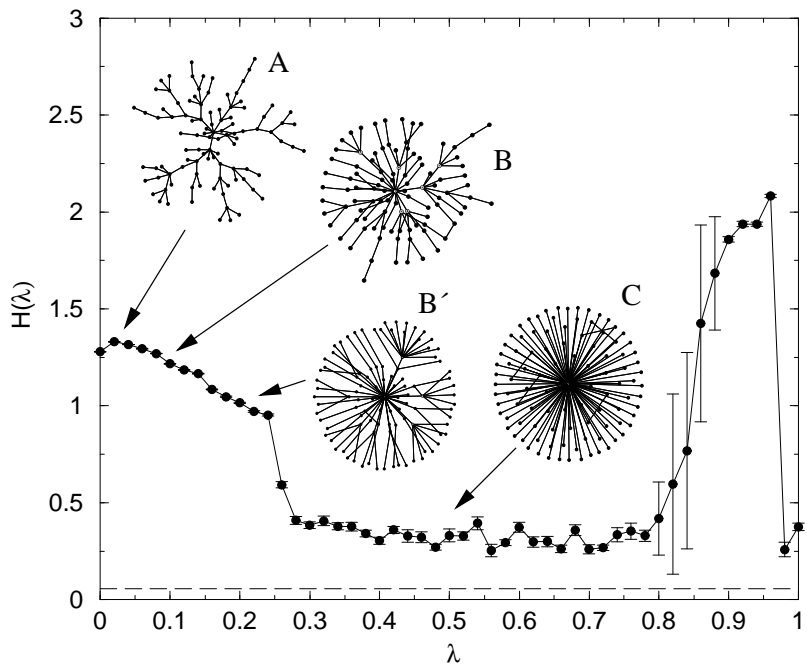}
\caption{\label{fig:entropy_versus_lambda} Average (over $50$ replicas)
 degree entropy as a function of $\lambda$ with $n=100$, $M={n \choose 2}$, 
$\mu=2/{n \choose 2}$ and $\rho(0)=0.2$.
Optimal networks for selected values of $\lambda$ are plotted. The entropy
 of a star network, $H_{star}=\log n - [(n-1)/n]\log (n-1)=0.056$ is provided
 as reference (dashed line). A: an exponential-like network with $\lambda=0.01$.
 B: A scale-free network with $\lambda=0.08$. 
Hubs involving multiple connections and a dominance 
of nodes with one connection can be seen. C: a star network with $\lambda=0.5$.
 B': a intermediate graph between B and C in which many hubs can be identified. 
}
\end{figure*}

For the sake of simplicity, we take an undirected graph having a fixed 
number of nodes $n$ and links defined by a binary adjacency matrix 
$A=\{a_{ij}\}$, $1\leq i,j \leq n$. Given a pair of 
vertices $i$ and $j$, $a_{ij}=1$ if they are linked ($a_{ij}=0$ otherwise) 
and $D_{ij}$ is the minimum distance 
between them. 
At time $t=0$, we have a randomly wired (Poissonian) graph in which two 
given nodes are connected with some probability $p$. The energy function of our 
optimization algorithm is defined as 
$$E(\lambda)=\lambda d+(1-\lambda)\rho$$ 
where $0 \leq \lambda,d,\rho \leq 1$ . $\lambda$ is a parameter 
controlling the linear combination of $d$ and $\rho$. The normalized 
number of links, $\rho$ is defined in terms of $a_{ij}$  as 
$$ \rho=\frac{1}{{n \choose 2}}\sum_{i<j} a_{ij}$$
and it is equivalent to Eq. \ref{equation:density}.
The normalized vertex-vertex distance, $d$, is defined as $d=D/D^{max}$ being
$$D=\frac{1}{{n \choose 2}}\sum_{i<j}D_{ij}$$
the average minimum vertex-vertex distance and $D^{max}=(n+1)/3$ the maximum value of 
$D$ that can be achieved by a connected network. 
A graph whose adjacency matrix satisfies
$a_{ij}=1$ if $|i-j|=1$ and $a_{ij}=0$ otherwise 
has the maximum average vertex-vertex distance that can be achieved 
by a connected graph of order $n$.

The minimization of $E(\lambda)$ involves the simultaneous minimization of distance and number 
of links (which is associated to cost). Notice that minimizing $E(\lambda)$  
implies connectedness (i.e. finite vertex-vertex distance) except for $\lambda=0$,
 where it will be explicitely enforced. 

The minimization algorithm proceeds as follows. At time $t=0$, the network is set up with
 a density $\rho(0)$ following a Poissonian distribution of degrees 
(connectedness is enforced). At time $t>0$, the graph is modified by randomly 
changing the state of some pairs of vertices. Specifically, with probability 
$\mu$, each $a_{ij}$ can 
switch from $0$ to $1$ or from $1$ to $0$. The new adjacency matrix is 
accepted if $E(\lambda,t+1)<E(\lambda,t)$. 
Otherwise, we try again with a different set of changes. The algorithm 
stops when the modifications on $A(t)$ are not accepted $M$ times in a row.
 Hereafter, $n=100$ \cite{NetworkOrderComment}, $M={n \choose 2}$ 
\cite{NetworkMaximumFailuresAllowedComment}, $2/{n \choose 2}$
 \cite{NetworkMutationRateCommnent} and $\rho(0)=0.2$.

We define the degree entropy on a certain value of $\lambda$ as
$$H=-\sum_{k=1}^{n-1} p_k \log p_k$$ 
where $p_k$ is the frequency
of vertices having degree $k$ and $\sum_{k=1}^{n-1} p_k=1$. 
This type of informational entropy will be used in our
characterization of the different phases \cite{EntropyMeasuresComment}. 

In Fig. \ref{fig:properties_versus_lambda} some of the basic average 
properties displayed by the optimized nets are shown against
$\lambda$. These plots, together with the degree entropy in 
Fig. \ref{fig:entropy_versus_lambda} suggest that four phases 
are present, separated by three sharp transitions at $\lambda_1^*
\approx 0.25$, $\lambda_2^* \approx 0.80$ and $\lambda_3^* \approx 0.95$ 
(see arrows in Fig. \ref{fig:properties_versus_lambda}). The
 second one separates sparse from dense nets and fluctuations
in $H(\lambda_3^*$) are specially high. $\rho(\lambda),C(\lambda) \approx 1$ for $\lambda>\lambda_3^* \approx 0.95$.
For $\lambda=0$ and $\lambda=1$ a Poissonian and a clique ($\rho(\lambda)=1)$ network are predicted, respectively. 

\begin{figure}
\includegraphics[scale=0.31]{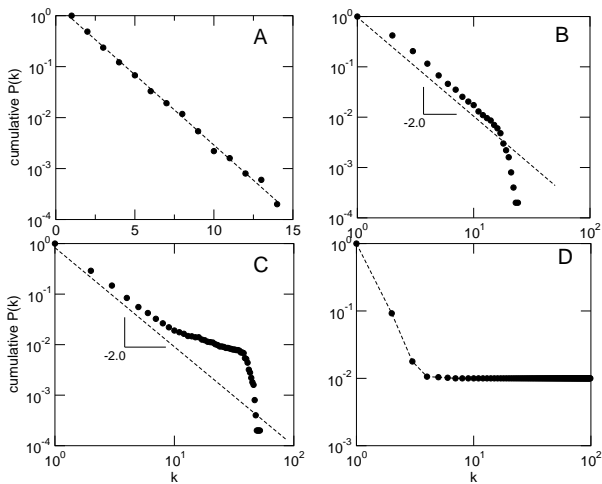}
\caption{\label{fig:degree_distributions} 
Selected cumulative degree distributions of networks obtained minimizing
 $E(\lambda$. Every distribution is an average over $50$ optimized networks 
with $n=100$, $M={n \choose 2}$, $\mu=2/{n \choose 2}$ and $\rho(0)=0.2$. A: 
an exponential-like distribution for $\lambda=0.01$. B: a power distribution 
with exponent $\gamma=2.0$ for $\lambda=0.08$ (with a sharp cutoff at
$\xi \approx 20$). C: $\lambda=0.20$. D: $\lambda=0.50$ (almost an star graph).}
\end{figure}

When taking a more careful look at the sparse domain $(0,\lambda_2^*)$, three 
non-trivial types of networks are obtained as $\lambda$ grows:  
\begin{itemize}
\item[a.]
Exponential networks, i. e. $P(k) \sim e^{-k/\xi}$. 
\item[b.]
Truncated scale-free networks, i. e. $P(k) \sim k^{-\gamma}
e^{-k/\xi}$ with $\gamma=3.0$ and $\xi \approx 20$ (for $n=100$).
\item[c.]
Star network phase ($\lambda_1^*< \lambda<\lambda_2^*$) 
i.e. a central vertex to which the rest of the vertices are 
connected to (no other connections are possible). Here, 
\begin{equation}
p_k=\frac{n-1}{n}\delta_{k,1}+\frac{1}{n}\delta_{k,n-1}
\end{equation}
A star graph has the shortest vertex-vertex distance 
between vertices among all the graphs having a minimal amount of links 
\cite{StarNetworkComment}. Non-minimal densities can be compensated with a 
decrease in distance, so pure star networks are not generally obtained.
\end{itemize}

The distributions of (a-c) types and that of a dense network are shown
 in Figure \ref{fig:degree_distributions}. A detailed examination of the 
transition between degree distributions reveals that hub formation explains
 the emergence of (b) from (a), hub competition (b') precedes the emergence of a 
central vertex in (c). The emergence of dense graphs from (c) consists
 of a progressive increase in the average degree of non-central vertices and a
 sudden loss of the central vertex. The transition to the star net
 phase is sharp.  
Fig. \ref{fig:entropy_versus_lambda} shows $\left<H(\lambda)\right>$
along with plots of the 
major types of networks. It can be seen that scale-free networks (b) are found
close to $\lambda_1^*$. The cumulative exponent 
of such scale-free networks is $\gamma=2.0$, the same that it would be expected for
 a random network generated with the Barab\'{a}si-Albert model \cite{Barabasi1999}.
 A necessary condition for the emergence of scaling in such model is the 
use of preferential attachment when connecting vertices. 

Our scenario suggests that
 preferential attachment networks might emerge at the boundary between
random attachment networks (a) and forced attachment (i.e. every vertex connected 
to a central vertex) networks (c) and points that optimization can
explain the selection of preferential attachment strategies in real complex networks.  
In our study, exponential-like distributions appear when distance is minimized under
 high density pressure, in agreement with the study by Amaral and
 co-workers on classes of small-world networks \cite{Amaral2000}. This
 might be the case of the power grid and of neural networks \cite{Amaral2000}. 

Whether or not optimization plays a key role in shaping the evolution
of complex networks, both natural and artificial, is an important
question. Different mechanisms have been suggested to explain the
emergence of the striking features displayed by complex networks. Most
mechanisms rely on preferential attachment-related rules, 
but other scenarios have also been suggested
\cite{Sole2001,Vazquez2001} in which external parameters have to be tuned. When
dealing with biological networks, the interplay between emergent
properties derived from network growth and selection pressures has to
be taken into account. As an example, metabolic networks seem to
result from an evolutionary history in which both preferential
attachment and optimization are present. 
The topology displayed by metabolic networks is scale-free, and the
underlying evolutionary history of these nets suggests that 
preferential attachment might have been involved \cite{Fell2001}. Early in the
evolution of life, metabolic nets grew by adding new metabolites, and
the most connected are actually known to be the oldest ones. On the 
other hand, several studies have revealed that metabolic pathways have
been optimized through evolution in a number of ways. This suggests
that the resulting networks are the outcome of both contributions,
plus some additional constraints imposed by the available components
to the evolving network \cite{Morowitz2001}. In this
sense, selective pressures might work by tuning underlying rules of 
net construction. This view corresponds to
Kauffman's suggestion that evolution would operate by taking advantage
of some robust, generic mechanisms of structure formation \cite{Kauffman1993}. 

\vspace{0.25 cm}
\begin{acknowledgments}
We acknowledge R. Pastor-Satorras for helpful discussions and the technical assistance
 of  F. Busquets. This work was supported by (and started at) 
the Santa Fe Institute (RFC and RVS) and grants of the Generalitat de 
 Catalunya (FI/2000-00393, RFC) and the CICYT (PB97-0693, RVS). 
\end{acknowledgments}


\begin{thebibliography}{30}
\bibitem[{Net({\natexlab{b}})}]{NetworkMaximumFailuresAllowedComment}
\bibinfo{note}{Intended for expecting that every pair of vertices has been
  allowed to change its state at least once.}
\bibitem[{Net({\natexlab{c}})}]{NetworkMutationRateCommnent}
\bibinfo{note}{We define the number of changes in the adjacency matrix between
  generations as $c= \left\{ a_{ij}(t+1)|i<j~and~a_{ij}(t+1) \neq a_{ij}(t)
  \right\}$. If $c=1$ then $d(t+1)<d(t)$ and $\rho(t+1) = \rho(t)$ is
  impossible. If $c>1$ then $d(t+1)<d(t)$ with $\rho(t+1)=\rho(t)$ is allowed.
  Thereafter $\mu$ is set to enforce $E[c]=2>1$.}
\bibitem[{Sta()}]{StarNetworkComment}
\bibinfo{note}{It can be shown through induction on $n$ that a star graph with
  a degree distribution \begin{equation}
  p_k=\frac{n-1}{n}\delta_{k,1}+\frac{1}{n}\delta_{k,n-1} \label{optimal_graph}
  \end{equation} has the minimum distance possible among all possible graphs
  having $n-1$ links. We define the distance of a graph of order $n$ as
  \begin{equation} d_n=\sum_{i<j} D_{ij} \end{equation} For $n=2$, the only
  connected graph (and thus the only with finite distance) trivially is the
  best one having $n-1$ links. If we assume that the graph described in Eq.
  \ref{optimal_graph} is the optimal for $n$ vertices, the optimal graph of
  $n+1$ vertices has $d_{n+1}=d_n+\Delta_{n+1}^k$ where $\Delta_{n+1}^k$ is the
  contribution to $D_{n+1}$ of the new vertex when linked to the $k$-th
  existing vertex. Thereafter, $\Delta_{n+1}^1=2n-1$ and
  $\Delta_{n+1}^k=3(n-1)$ for $1 < k \leq n$.
  $\Delta_{n+1}^1<\Delta_{n+1}^{k>1}$ holds for $n>2$, so the graph of order
  $n+1$ obeying Eq. \ref{optimal_graph} is also the best one with $n-1$ links.}
\end{thebibliography}
\end{document}